\documentclass[12pt]{article}
\usepackage{amssymb,amsmath,bm,epsfig,comment}

\renewcommand{\thefootnote}{\fnsymbol{footnote}}
\begin{document}
\title{}

\title{
\begin{flushright}
\begin{minipage}{0.2\linewidth}
\normalsize
WU-HEP-14-04 \\*[50pt]
\end{minipage}
\end{flushright}
{\Large \bf 
The Higgs boson mass and SUSY spectra in 10D SYM theory with magnetized 
extra dimensions\\*[20pt] } }

\author{
Hiroyuki~Abe\footnote{
E-mail address: abe@waseda.jp}, \ 
Junichiro~Kawamura\footnote{
E-mail address: junichiro-k@ruri.waseda.jp} \ and \
Keigo~Sumita\footnote{
E-mail address: k.sumita@moegi.waseda.jp}\\*[20pt]
{\it \normalsize 
Department of Physics, Waseda University, 
Tokyo 169-8555, Japan} \\*[50pt]}

\date{
\centerline{\small \bf Abstract}
\begin{minipage}{0.9\linewidth}
\medskip 
\medskip 
\small 
We study the Higgs boson mass and the spectrum of supersymmetric (SUSY) 
particles in the well-motivated particle physics model derived from 
a ten-dimensional supersymmetric Yang-Mills theory compactified on three 
factorizable tori with 
magnetic fluxes. This model was proposed in a previous work, 
where the flavor structures of the 
standard model including the realistic Yukawa hierarchies 
are obtained from non-hierarchical input parameters 
on the magnetized background. 
Assuming moduli- and anomaly-mediated contributions dominate 
the soft SUSY breaking terms, we study the precise SUSY spectra and analyze 
the Higgs boson mass in this mode, 
which are compared with the latest experimental data. 
\end{minipage}}

\begin{titlepage}
\maketitle
\thispagestyle{empty}
\clearpage
\tableofcontents
\thispagestyle{empty}
\end{titlepage}

\renewcommand{\thefootnote}{\arabic{footnote}}
\setcounter{footnote}{0}

\section{Introduction}

The discovery of a Higgs boson at the Large Hadron Collider (LHC) 
\cite{Aad:2012tfa,Chatrchyan:2012ufa} 
provides the last complement of the standard model (SM), 
and it is getting recognized as the successful and reliable theory 
more than before. 
However, various phenomenological (or other) facts 
clearly indicate the presence of new physics beyond the SM. 
Supersymmetry (SUSY) and extra dimensional space are 
the well-known great candidates for that and 
diverse models has been constructed with them. 
Now, the discovery of the Higgs boson enables us to further study or verify such models 
including the SUSY spectra and dynamics of the extra dimensional space.

The complicated structures, e.g., 
product gauge groups, three generations and hierarchical Yukawa couplings, 
are some of the most well-known mysteries of the SM. 
In this paper with the motive, we consider 
higher-dimensional supersymmetric Yang-Mills (SYM) theories 
compactified on simple factorizable tori with magnetic fluxes. 
The magnetic fluxes on the tori induce the four-dimensional 
(4D) chiral spectra 
\cite{Bachas:1995ik,Cremades:2004wa}, and furthermore, 
that is capable of originating the flavors of the SM. 
The degenerate zero-modes are induced by the magnetic fluxes with a certain 
degeneracy corresponding to the flux magnitude which is quantized due to 
the Dirac's argument, and their wavefunctions are localized 
at different points of the magnetized tori. 
As a result, Yukawa coupling constants in the 4D effective theory, 
which are given by overlap integrals of such localized wavefunctions, 
can be hierarchical among the degenerate zero-modes, that are the generations.

Toroidal compactifications with magnetic fluxes are quite attractive to derive 
the SM from the higher-dimensional SYM theories. 
The SUSY theories in higher-dimensional spacetime have 
$\mathcal N=2,3~{\rm or}~4$ SUSY counted by the 4D supercharges depending 
on the structure of 
background as well as the dimensionality of spacetime. 
The magnetic fluxes generally break the SUSY and careful analyses are required 
to construct MSSM-like models preserving the $\mathcal N=1$ SUSY, 
because the configuration of fluxes determine not only the number of SUSY 
preserved but also the almost everything mentioned above. 
In the 4D $\mathcal N=1$ superspace, from such a perspective, 
we proposed a systematic way of dimensional reduction of the ten-dimensional 
(10D) SYM theories compactified on three factorizable tori with magnetic 
fluxes \cite{Abe:2012ya}. 
Thanks to that, a MSSM-like model preserving the $\mathcal N=1$ SUSY was 
constructed \cite{Abe:2012fj}, where the magnetic fluxes originated the flavor 
structures of the SM particles and even of their superpartners by 
assuming some typical mediation mechanisms of SUSY breaking. 
Then the observational consistencies with respect to the SUSY flavor 
violations were also studied in Ref~\cite{Abe:2012fj}.

In this paper, we study the low-energy phenomenology of the model proposed by 
Ref~\cite{Abe:2012fj}, especially focusing on the Higgs boson mass 
and the SUSY spectrum precisely with the latest experimental data, 
and those two issues greatly correlate to each other. 
In the model, the background fluxes are already fixed to realize the SM flavor structures, 
then there are only a few remaining parameters 
which appear mainly from the degrees of freedom governing the (SUSY) dynamics 
of extra dimensional space, i.e., the moduli superfields, determining 
the SUSY spectrum in the MSSM sector. 
We will study the parameter dependence of the Higgs boson mass via the SUSY spectrum.

The organization of this paper is as follows. 
In Section \ref{sec:review}, we give a brief review of the magnetized model 
proposed in Ref.~\cite{Abe:2012fj} including the way of dimensional reduction 
with the superfield description \cite{Abe:2012ya}. 
In Section \ref{sec:yukawa}, we estimate the masses and mixing angles of the 
quarks and the leptons. 
We can calculate the Yukawa coupling constants depending on some parameters 
in the model. 
Such estimations have already been done, but we improve them especially 
taking the Higgs boson mass into account. 
On the background, we study the Higgs boson mass and the SUSY spectrum in 
Section \ref{sec:higgs}. 
We assume that the moduli- and anomaly-mediated contributions dominate 
the soft SUSY breaking terms in the MSSM sector, which are typical mediators 
in higher-dimensional spacetime, and parameterize them by the auxiliary 
$F$-components of the corresponding moduli and the compensator supermultiplets, respectively. 
Accordingly, we calculate the soft SUSY breaking parameters, 
the SUSY spectrum, and furthermore, the Higgs boson mass. 
Finally, Section \ref{sec:conc} is devoted to conclusions and discussions.

\section{Review of the magnetized model}\label{sec:review}
We give a review of the phenomenological model proposed in Ref.~\cite{Abe:2012fj}. 
The model was derived starting from the 10D $U(8)$ SYM theory compactified 
on a (factorizable) product of 4D Minkowski space and 
three two-dimensional (2D) tori, $R^{1,3}\times \prod_{i=1}^3 (T^2)_i$. 
The action is given by 
\begin{equation}
S=\int d^{10}x \sqrt {-G} \left\{-\frac1{4g^2}{\rm tr}\left(F^{MN}F_{MN}\right)
+\frac i{2g^2}{\rm tr}\left(\bar\lambda \Gamma^MD_M\lambda\right)\right\},\label{eq:sym}
\end{equation}
that contains the 10D vector field $A_M$ ($M = 0,~1,~\cdots,~9$) 
and the 10D Majorana-Weyl spinor field $\lambda$. 
For each $i=1,2,3$, the metric of 2D torus $(T^2)_i$ is expressed as 
\begin{equation*}
g^{(i)}= \left(2\pi R^{(i)}\right)^2
\begin{pmatrix}
1&{\rm Re}\tau^{(i)}\\
{\rm Re}\tau^{(i)}&\left|\tau^{(i)}\right|^2
\end{pmatrix},
\end{equation*}
where the real parameters $R^{(i)}$ correspond to typical sizes of the three torus 
and the complex parameters $\tau^{(i)}$ to the complex structures 
($i=1,~2,~3$), and those are all contained in the 10D metric $G_{MN}$ whose 
determinant is described by $G$ in the action. 

We will introduce magnetic fluxes on each of the factorizable tori and derive the 4D effective action. We have an extremely useful superfield description for such dimensional reductions. 
(The details are given in Ref.~\cite{Abe:2012ya}.) 
First, we define the complex coordinates $z_i$ 
of the $i$-th 2D torus $(T^2)_i$ as 
\begin{equation*}
z^i\equiv\frac12\left(x^{2+2i}+\tau^{(i)}x^{3+2i}\right),
\end{equation*}
and its metric $h_{i\bar j}=2(2\pi R^{(i)})^2\delta_{i\bar j}=
\delta_{\rm i\bar j}e_i^{~\rm i}e_{\bar j}^{~\rm \bar j}$ 
with which the line element in 6D extra space is given by 
$ds^2_{\rm 6D}=2h_{i\bar j}dz^id\bar z^{\bar j}$ where 
$\bar z^{\bar i}$ is the complex conjugate to $z^i$. 
The 10D vector field $A_M$ is decomposed into the 4D vector field $A_\mu$ 
and the others, and we define three complex fields, 
\begin{equation*}
A_i\equiv -\frac1{{\rm Im}\tau^{(i)}}\left({\bar\tau^{(i)}}A_{2+2i}-A_{3+2i}\right).
\end{equation*}
The 10D Majorana-Weyl spinor field $\lambda$ is also decomposed into 
four 4D Weyl spinors, 
$\lambda_0=\lambda_{+++}$, $\lambda_1=\lambda_{+--}$, 
$\lambda_2=\lambda_{-+-}$ and $\lambda_3=\lambda_{--+}$. 
The subscripts $\pm$ represent the chirality on the three tori 
and other combinations of them , $\lambda_{---}$, $\lambda_{-++}$, 
$\lambda_{+-+}$ and $\lambda_{++-}$ do not appear 
because of the 10D Majorana-Weyl condition. 

These form a vector multiplet $\left\{A_\mu,~\lambda_0\right\}$ and 
three chiral multiplets $\left\{A_i,~\lambda_i\right\}$ under 
a 4D $\mathcal N=1$ SUSY, which are assigned to a 4D $\mathcal N=1$ 
vector superfield $V$ and three chiral superfields $\phi_i$ 
respectively as follows, 
\begin{eqnarray}
V & \equiv & -\theta\sigma^\mu\bar\theta A_\mu +i\bar\theta\bar\theta\theta\lambda_0
-i\theta\theta\bar\theta\bar\lambda_0+\frac12\theta\theta\bar\theta\bar\theta D,\nonumber\\
\phi_i & \equiv & \frac1{\sqrt2} A_i+\sqrt2\theta\lambda_i+\theta\theta F_i,\label{eq:phii}
\end{eqnarray} 
where the $\theta$ denotes the Grassmann coordinate of the 
$\mathcal N=1$ superspace. 
These superfields lead us to rewrite the SYM action (\ref{eq:sym}) 
in the 4D $\mathcal N=1$ superspace as \cite{Abe:2012ya, Marcus:1983wb} 
\begin{equation}
S=\int d^{10}x\sqrt {-G}\left[\int d^4\theta \mathcal K +\left\{
\int d^2\theta\left(\frac1{4g^2}\mathcal W^\alpha\mathcal W_\alpha\right)
+{\rm h.c.}\right\}\right],\label{eq:10dsym1}
\end{equation}
with the following functions of superfields, 
\begin{eqnarray}
\mathcal K &=& \frac2{g^2}h^{i\bar j}{\rm Tr}\left[
\left(\sqrt2\bar\partial_{\bar i}+\bar\phi_{\bar i}\right)e^{-V}
\left(-\sqrt2\partial_j+\phi_j\right)e^V
+\bar\partial_{\bar i}e^{-V}\partial_je^V\right]+\mathcal K_{\rm WZW},\nonumber\\
\mathcal W &=& \frac1{g^2}\epsilon^{\rm ijk}e_{\rm i}^{~i}e_{\rm j}^{~j}
e_{\rm k}^{~k} {\rm tr}\left[\sqrt2\phi_i\left(\partial_j\phi_k-\frac1{3\sqrt2}
\left[\phi_j, \phi_k\right]\right)\right],\nonumber\\
\mathcal W_\alpha &=& -\frac14\bar D\bar De^{-V}D_\alpha e^V,\label{eq:10dsym2}
\end{eqnarray}
where $\partial_i$ is a derivative with respect to the complex 
coordinate $z_i$, and $h_{ij}$ and $e_i^{~\rm i}$ are the metric and vielbein 
of the three tori. The terms $\mathcal K_{\rm WZW}$ vanish 
in the Wess-Zumino gauge. 
In the definition of $\mathcal W_\alpha$, 
$D_\alpha$ and $\bar D_{\alpha}$ denote the (super)covariant  
derivatives. The 10D SYM action with the superfield description has 
the $\mathcal N=4$ SUSY, and a $\mathcal N=1$ SUSY, which is a part of that, 
becomes manifest. 
The superfields $V$ and $\phi_i$ contain the auxiliary fields $D$ and $F_i$ 
respectively, 
and it is obvious that the $\mathcal N=1$ SUSY is preserved if their vacuum expectation values (VEVs) vanish.

We consider the following magnetized background, 
\begin{equation}
\langle A_i\rangle =\frac{\pi}{{\rm Im} \tau^{(i)}}
\left(M^{(i)}\bar z_{\bar i}+\bar \zeta_i\right), \label{eq:oneflux}
\end{equation} 
where $M^{(i)}$ and $\zeta_i$ are matrices of the number of magnetic fluxes 
and of the Wilson-lines 
on $(T^2)_i$. 
The SUSY preserving condition $\langle D\rangle=0$ requires 
(The $(1,1)$ form fluxes (\ref{eq:oneflux}) automatically satisfy 
$\langle F_i\rangle=0$.) 
\begin{equation} 
\frac1{\mathcal A^{(1)}}M^{(1)}+\frac1{\mathcal A^{(2)}}M^{(2)}
+\frac1{\mathcal A^{(3)}}M^{(3)}=0, \label{eq:susycond}
\end{equation} 
where $\mathcal A^{(i)}=(2\pi R^{(i)})^2\,{\rm Im}\,\tau^{(i)}$ expresses 
the area of $(T^2)_i$. 
In our model, the magnetic fluxes and Wilson-line matrices 
are encoded in the following $(8\times8)$ diagonal matrices whose rows and 
columns cover the space of the U(8) gauge group, 
\begin{equation}
M^{(i)}=\begin{pmatrix}
M_C^{(i)}{\bm 1}_4&0&0\\
0&M_L^{(i)}{\bm 1}_2&0\\
0&0&M_R^{(i)}{\bm 1}_2
\end{pmatrix},\hspace{15pt}
\zeta^{(i)}=\begin{pmatrix}
\zeta_C^{(i)}{\bm 1}_3&0&0&0&0\\
0&\zeta_{C'}^{(i)}&0&0&0\\
0&0&\zeta_L^{(i)}{\bm 1}_2&0&0\\
0&0&0&\zeta_{R'}^{(i)}&0\\
0&0&0&0&\zeta_{R''}^{(i)}
\end{pmatrix}.\label{eq:magwil}
\end{equation} 
The explicit values of $M_a^{(i)}$ and 
$\zeta_a^{(i)}$ are determined phenomenologically which are shown later 
for each $a=C,C',L,R',R''$. 
The flux matrices $M^{(i)}$ must have the three-block diagonal forms 
because that is strongly constrained by the SUSY condition 
and the diagonal components $M^{(i)}_a$ are integers due to the Dirac's 
quantization condition. 
They break the $U(8)$ gauge group down to 
$U(4)_C\times U(2)_L\times U(2)_{R}$, the Pati-Salam gauge group, 
and that is further broken down by the Wilson-lines 
close to the SM gauge group. 
There are five unbroken $U(1)$s and a linear combination of them 
will be the $U(1)$ hypercharge. 
The gauge bosons of the other four $U(1)$s than that of the $U(1)_Y$ 
are assumed to become massive due to some UV physics 
(e.g., the Green-Schwartz mechanism~\cite{Green:1984sg}) 
and decouple from the physics below the compactification scale. 
We use the indices, $a,b=C,C',L,R',R''$ for these remaining subgroups 
and $\phi_i^{ab}$ represents a bifundamental representation 
$(N_a, \bar N_b)$ of $U(N_a)\times U(N_b)$, included in $\phi_i$ 
which is the adjoint representation of $U(8)$ for each $i=1,2,3$. 
These notations are similarly adopted for $V$ which is also the adjoint 
representation of $U(8)$.

From now on, we are focusing on only the zero-modes of $V^{ab}$ and 
$\phi_i^{ab}$ to construct the phenomenological model 
which is being implicit, and then we use the same notation for the zero-mode 
as the one for the corresponding 10D fields. 
The degenerate zero-modes appear depending on the flux configuration and their 
wavefunctions can be obtained solving the zero-mode equations 
\cite{Cremades:2004wa, Abe:2012ya}. In the following, we summarize the results 
briefly.

First, the diagonal parts of the vector superfield $V^{aa}$ correspond to the 
SM gauge fields. They feel no magnetic flux and have a flat wavefunction 
on the torus. 
Those of the chiral superfields $\phi_i^{aa}$ also remain as massless 
vector-like exotics or notorious open string moduli. 
In the phenomenological model building, some prescriptions are required 
to treat these zero-modes $\phi_i^{aa}$ if exist as we will mention it later.

Next, the off-diagonals $V^{ab} (a\neq b)$ have heavy masses 
in response to the partial breaking of the $U(8)$ gauge group, and they 
have no effect on the phenomenologies at the low-energy below 
the compactification scale. 
As for the bifundamentals $\phi_i^{ab}$, they carry the matter fields 
and the Higgs fields of the MSSM, which are the most important 
phenomenological ingredients. 
The bifundamentals $\phi_i^{ab}$ feel the magnetic fluxes 
$M^{(j)}_{ab}\equiv M^{(j)}_{a}-M^{(j)}_{b}$ on the torus $(T^2)_j$. 
In the case with $i=j$,  degenerate zero-modes arise if and only if 
$M^{(j)}_{ab}>0$, whose degeneracy is also given by $M^{(j)}_{ab}$, 
while the conjugate representations $\phi_i^{ba}$ are eliminated 
because they feel the negative magnetic fluxes $M^{(j)}_{ba}<0$. 
On the other hand, in the case with $i\neq j$, negative magnetic 
fluxes $M^{(j)}_{ab}<0$ yield the degenerate zero-modes 
with the degeneracy $|M^{(j)}_{ab}|$, 
and their conjugates are projected out because of the positive magnetic 
fluxes $M^{(j)}_{ab}>0$. 
Consequently, the magnetic fluxes cause a kind of projection 
generating a 4D chiral spectra with the degeneracies identified 
as generations. 
If there are vanishing magnetic fluxes $M^{(j)}_{ab}=M^{(j)}_{ba}=0$, 
the projection does not occur and both of the single zero-modes 
$\phi_i^{ab}$ and $\phi_i^{ba}$ remain simultaneously with flat wavefunctions 
regardless of whether $i=j$ or not. 
We can identify the degenerate zero-modes as the generations 
of the MSSM matters.

Furthermore, the magnetic fluxes give the Gaussian-like profiles to 
the zero-mode wavefunctions, and each degenerate zero-mode is localized 
at the different point on the magnetized torus from each other. 
The Yukawa coupling constants in the 4D effective action are given by 
overlap integrals of these Gaussian-like wavefunctions. 
The magnitude of the overlaps determines the values of the Yukawa coupling constants, 
while the Wilson-lines can shift the peak position of each Gaussian 
differently in general, that is, 
they can control the value of the Yukawa coupling constants hierarchically.

Now, we show the phenomenologically specific flux configuration, 
\begin{eqnarray}
\left(M_C^{(1)},~M_L^{(1)},~M_R^{(1)}\right)&=&\left(0,~3,~-3\right),\nonumber\\
\left(M_C^{(2)},~M_L^{(2)},~M_R^{(2)}\right)&=&\left(0,~-1,~0\right),\nonumber\\
\left(M_C^{(3)},~M_L^{(3)},~M_R^{(3)}\right)&=&\left(0,~0,~1\right), \label{eq:magconfig} 
\end{eqnarray} 
on which magnetized background, the SUSY preserving condition (\ref{eq:susycond}) 
is satisfied if 
\begin{equation} 
\mathcal A^{(1)}/\mathcal A^{(2)} = \mathcal A^{(1)}/\mathcal A^{(3)}=3. 
\label{eq:arear}
\end{equation}
This flux configuration is the unique one to produce the 
three-generation MSSM-like 
model preserving the $\mathcal N=1$ SUSY unless 
we consider more complicated magnetized backgrounds than 
Eq.~(\ref{eq:oneflux}), see Ref.~\cite{Abe:2013bba}. 
On the background (\ref{eq:magconfig}), the three phenomenologically relevant 
sectors feel the magnetic fluxes as shown in Table~\ref{tb:model336}. 
\begin{table}[htb]
\begin{center}
\begin{tabular}{c|c|c|c}\hline
&Left-handed&Higgs&Right-handed\\ 
&$M^{(i)}_C-M^{(i)}_L$&$M^{(i)}_L-M^{(i)}_R$&$M^{(i)}_R-M^{(i)}_C$\\ \hline
$T^2_1$ &$-3$&$6$&$-3$\\
$T^2_2$ &$+1$&$-1$&$0$\\
$T^2_3$ &$0$&$-1$&$+1$\\ \hline
\end{tabular}
\end{center}
\caption{The number of magnetic fluxes felt by each sector on each torus.}
\label{tb:model336}
\end{table}
From the table, we see the magnetic fluxes realize the three generations 
of quark and 
lepton multiplets, and the six generations of Higgs multiplets. 
There also remain some extra fields, e.g., massless exotics and 
open string moduli as mentioned above, but we can eliminate most of them 
with a certain orbifold projection without any changes in the MSSM sector. 
For such a purpose, we consider a $Z_2$ orbifold defined on two tori $(T^2)_2$ 
and $(T^2)_3$ as
\begin{equation*}
z_1\rightarrow z_1,~~~z_2\rightarrow -z_2,~~~z_3\rightarrow -z_3, 
\end{equation*}
with the $Z_2$ twists acting on the fields of the 10D SYM theory as 
\begin{eqnarray*}
V(x_\mu, z_1, z_2, z_3) &=& +P V(x_\mu, z_1, -z_2, -z_3) P,\\
\phi_1(x_\mu, z_1, z_2, z_3) &=& +P \phi_1(x_\mu, z_1, -z_2, -z_3) P,\\
\phi_2(x_\mu, z_1, z_2, z_3) &=& -P \phi_2(x_\mu, z_1, -z_2, -z_3) P,\\
\phi_3(x_\mu, z_1, z_2, z_3) &=& -P \phi_3(x_\mu, z_1, -z_2, -z_3) P,
\end{eqnarray*}
where $P$ is an $(8\times8)$ projection operator ($P^2={\bm 1}$) written 
in the following form, 
\begin{equation*}
P=\begin{pmatrix}
-{\bm 1}_4&0&0\\
0&+{\bm 1}_2&0\\
0&0&+{\bm 1}_2
\end{pmatrix}.
\end{equation*} 
After this $Z_2$ projection, 
the remaining zero-mode contents can be expressed as follows, 
\begin{align*}
\phi_1^{{\cal I}_{ab}} &= 
\left( 
\begin{array}{cc|c|cc}
\Omega_C^{(1)}  & \Xi_{CC'}^{(1)} & 0 &0&0\\
\Xi_{C'C}^{(1)} & \Omega_{C'}^{(1)} & 0 &0&0\\
\hline 
0 &0 & \Omega_L^{(1)} & H_u^K & H_d^K \\ 
\hline 
0 & 0 & 0 & \Omega_{R'}^{(1)} & \Xi_{R'R''}^{(1)} \\
0 & 0 & 0 & \Xi_{R''R'}^{(1)} & \Omega_{R''}^{(1)} 
\end{array}
\right), 
\nonumber \\
\phi_2^{{\cal I}_{ab}} &= 
\left( 
\begin{array}{cc|c|cc}
0 & 0 & Q^I & 0 & 0 \\
0& 0& L^I & 0 & 0 \\
\hline 
0 & 0 & 0& 0 & 0 \\
\hline 
0 & 0 & 0 & 0 & 0 \\
0 & 0 & 0 & 0 &0 
\end{array}
\right), ~~~~~
\phi_3^{{\cal I}_{ab}} = 
\left( 
\begin{array}{cc|c|cc}
0&0  & 0 & 0 & 0 \\
0& 0 & 0 & 0 & 0 \\
\hline
0 & 0 & 0& 0 & 0 \\
\hline 
U^J & N^J & 0 &0 &0 \\
D^J & E^J & 0 & 0 & 0
\end{array}
\right), 
\end{align*}
where $5 \times 5$ block submatrices are matrix representations of 
the remaining gauge subgroups similar to Eq.~(\ref{eq:magwil}). 
Indices $I,J=1,2,3$ label the degenerate zero-modes of the matter sector, 
that is the three generations of quarks and leptons. 
The Higgs multiplets also have an index 
$K=1,~2,\cdots~6$. 
Only a few of extra fields survive, $\Omega$ and $\Xi$. 
Again phenomenologically a certain prescription would be required to 
give them heavy masses somehow to be decoupled from the MSSM, 
which is beyond the scope of this paper.

Under the above $Z_2$ twist, nonvanishing Wilson-lines on tori 
$(T^2)_2$ and $(T^2)_3$, 
$\zeta^{(2)}_a, \zeta^{(3)}_a\neq0$, are forbidden. 
However, only $\zeta^{(1)}_a$ is effective for the flavor structure 
because the magnetic fluxes shown in Eq.~(\ref{eq:magconfig}) 
cause the three-generation structure solely  on torus $(T^2)_1$. 
Therefore only the Wilson-line parameters $\zeta^{(1)}_a$ are enough to 
control the hierarchical structure of the Yukawa matrices in the MSSM sector. 
We will assign specific values to $\zeta^{(1)}_a$ in the next section.

On this non-trivial background, we can derive the 4D effective action 
of the 10D SYM theory with the superfield description 
(\ref{eq:10dsym1}) and (\ref{eq:10dsym2}). 
It contains the gauge and matter kinetic terms, gauge interaction terms and 
Yukawa coupling terms of the MSSM, 
and the coefficients of them are given as functions of 
the 10D gauge coupling constant $g$, the torus radii $R^{(i)}$ and 
the torus complex structures $\tau^{(i)}$. 
The values of them will be given as VEVs of moduli fields in supergravity 
(SUGRA). 
The moduli supermultiplets on the magnetized tori
consist of dilaton, K\"ahler moduli and complex structure moduli 
chiral supermultiplets\footnote{
The definitions of moduli multiplets are the same as 
those for pure factorizable tori regardless of the existence of YM fluxes. }, 
which are denoted as superfields $S$, $T_i$ and $U_i$ respectively. 
We parameterize their VEVs as 
\begin{equation*}
\langle S\rangle = s +\theta\theta F^s,\hspace{20pt}
\langle T_i\rangle = t_i +\theta\theta F^{t_i},\hspace{20pt}
\langle U_i\rangle = u_i +\theta\theta F^{u_i}, 
\end{equation*}
and the relations between $\left\{s,~t_i,~u_i\right\}$ 
and $\left\{g,~R^{(i)},~\tau^{(i)}\right\}$ are given by 
\begin{equation*}
{\rm Re }\, s = g^{-2}\mathcal A^{(1)}\mathcal A^{(2)}\mathcal A^{(3)},\hspace{20pt}
{\rm Re }\, t_i = g^{-2}\mathcal A^{(i)},\hspace{20pt}
u_i = i\, \bar \tau^{(i)}. 
\end{equation*}
Each $F^m$ of the supermultiplets $m=s,t_i,u_i$ is a parameter describing 
the magnitude of the SUSY breaking 
mediated by each moduli chiral multiplet. 
Using these relations, we can determine the moduli dependence 
of the 4D effective action and construct a 4D effective SUGRA action. 
The general action of the 4D $\mathcal N=1$ effective SUGRA 
(on the 4D gravitational background) 
is given in the superspace as 
\begin{equation*}
S = \int\, d^4x \sqrt {-g^C} \left[-3\int\, d^4\theta CC^*
\left(e^{K/3}\right)+\left\{
\int\,d^2\theta 
\left(\frac14 f_aW^{a\alpha}W^a_{\alpha} +C^3 W +\right)+h.c. 
\right\} \right],
\end{equation*} 
using a chiral compensator superfield $C$ whose $F$-components will be 
denoted by $F^C$ later. 
Finally, we can find the MSSM sector there, where the 
K\"ahler and superpotential, $K$ and $W$, can be expanded as 
\begin{eqnarray*}
K &=& K_0 + Z_{I\bar I}^{(Q_L)}Q_L^I\bar Q_L^{\bar I} 
+ Z_{J\bar J}^{(Q_R)}Q_R^J\bar Q_R^{\bar J}
+ Z_{K\bar K}^{(H)}H^K\bar H^{\bar K}+\cdots,\\
W &=& W_0 +  \lambda^{(Q_R)}_{IJK}Q_L^IQ_R^JH^K +\cdots. 
\end{eqnarray*}
The chiral superfields $Q_L^I = \left\{Q^I,~L^I\right\}$, 
$Q_R^J = \left\{U^J,~D^J,~N^J,~E^J\right\}$ 
and $H^K = \left\{H_u^K,~H_d^K\right\}$ as well as 
(implicit) vector superfields in the MSSM appear with the 
K\"ahler metrics $Z_{I\bar I}^{(Q_L)}, Z_{J\bar J}^{(Q_R)}$ and 
$Z_{K\bar K}^{(H)}$, the holomorphic Yukawa couplings 
$\lambda^{Q_R}_{IJK}$ as well as the gauge kinetic functions $f_a$ 
as functions of the seven moduli chiral superfields. 
The K\"ahler metrics are given by 
\begin{eqnarray}
Z^{(Q_L)}_{I\bar I} &=& \delta_{I\bar I}\frac1{\sqrt3} \left(T_2+\bar T_2\right)^{-1}\left(U_1+\bar U_1\right)^{-1/2}
\left(U_2+\bar U_2\right)^{-1/2}\exp
\frac{4\pi\left({\rm Im}\,\zeta_{Q_L}\right)^2}{3\left(U_1+\bar U_1\right)},\nonumber\\
Z^{(Q_R)}_{J\bar J} &=& \delta_{J\bar J}\frac1{\sqrt3} \left(T_3+\bar T_3\right)^{-1}\left(U_1+\bar U_1\right)^{-1/2}
\left(U_3+\bar U_3\right)^{-1/2}\exp
\frac{4\pi\left({\rm Im}\,\zeta_{Q_R}\right)^2}{3\left(U_1+\bar U_1\right)},\nonumber\\
Z^{(H)} _{K\bar K}&=& \delta_{K\bar K}\sqrt6 \left(T_1+\bar T_1\right)^{-1}
\left\{\prod_{i=1}^3\left(U_i+\bar U_i\right)^{-1/2}\right\}\exp
\frac{-4\pi\left({\rm Im}\,\zeta_{H}\right)^2}{6\left(U_1+\bar U_1\right)},
\label{eq:zzz}
\end{eqnarray}
where the Wilson-lines $\zeta_{Q_L,Q_R,H}$ distinguish the up/down sectors of 
quark/lepton multiplets, 
but they are universal for the generations, which are expressed as 
\begin{align*}
\zeta_Q&\equiv\zeta^{(1)}_C-\zeta^{(1)}_L,&~~~
\zeta_L&\equiv\zeta^{(1)}_{C'}-\zeta^{(1)}_L,&~~~
\zeta_{H_u}&\equiv\zeta^{(1)}_L-\zeta^{(1)}_{R'},&~~~
\zeta_{H_d}&\equiv\zeta^{(1)}_L-\zeta^{(1)}_{R''},\\
\zeta_U&\equiv\zeta^{(1)}_{R'}-\zeta^{(1)}_C,&~~~
\zeta_D&\equiv\zeta^{(1)}_{R''}-\zeta^{(1)}_C,&~~~
\zeta_N&\equiv\zeta^{(1)}_{R'}-\zeta^{(1)}_{C'},&~~~
\zeta_E&\equiv\zeta^{(1)}_{R''}-\zeta^{(1)}_{C'}.
\end{align*} 
The holomorphic Yukawa couplings are written as 
\begin{equation}
\lambda_{IJK}^{(Q_R)} = \sum_{m=1}^6\delta_{I+J+3\left(m-1\right),K}\,
\vartheta\begin{bmatrix}
\frac{3\left(I-J\right)+9\left(m-1\right)}{54}\\0
\end{bmatrix}
\left(3\left(\bar\zeta_{Q_L}-\bar\zeta_{Q_R}\right), 54 i U_1\right), \label{eq:lambda}
\end{equation}
where the explicit generation dependences arise as a consequence of the 
localized zero-modes, and the Wilson lines distinguish 
the right-handed sectors $Q_R$. $\vartheta$ is a Jacobi-theta function. 
Finally, the gauge kinetic functions $f_a = S$ are universal for 
all the gauge groups. 
We have successfully obtained the 4D effective SUGRA action focusing 
on the MSSM contents.

\section{Numerical analysis of Yukawa matrices}\label{sec:yukawa} 
In this section, we estimate the masses and mixing angles of the 
quarks and leptons numerically. 
We have the six generations of Higgs multiplets and identify 
one linear combination of them 
as the MSSM Higgs multiplets. 
Then the physical Yukawa matrices are expressed as 
\begin{align*}
y^u_{IJ}v_u&=\frac{\lambda^{(U)}_{IJK}\langle H_u^K\rangle}{\sqrt{e^{-K_0}Z^{(Q)}Z^{(U)}Z^{(H_u)}}},&
\hspace{5pt}
y^d_{IJ}v_d &= \frac{\lambda^{(D)}_{IJK}\langle H_d^K\rangle}{\sqrt{e^{-K_0}Z^{(Q)}Z^{(D)}Z^{(H_d)}}},
\nonumber\\
y^{\nu}_{IJ}v_u&=\frac{\lambda^{(N)}_{IJK}\langle H_u^K\rangle}{\sqrt{e^{-K_0}Z^{(L)}Z^{(N)}Z^{(H_u)}}},&
\hspace{5pt}
y^e_{IJ}v_d &= \frac{\lambda^{(E)}_{IJK}\langle H_d^K\rangle}{\sqrt{e^{-K_0}Z^{(L)}Z^{(E)}Z^{(H_d)}}},
\end{align*}
where $v_u$ and $v_d$ are the VEVs of up- and down-type Higgs fields 
in the MSSM respectively, 
and we note that there is a summation over the generations 
$K=1,2,\cdots,6$ of Higgs fields in each matrix. 
The denominators appear as a consequence of the canonical normalizations 
of fields. The moduli K\"ahler potential $K_0$ of the three factorizable tori 
$\prod_{i=1}^3 (T^2)_i$ is given by 
\begin{equation*}
K_0 = -\ln \left(S+\bar S\right) -\sum_i \ln \left(T_i+\bar T_i\right) 
-\sum_i \ln \left(U_i+\bar U_i\right). 
\end{equation*}
Now we can estimate 
the values of the Yukawa coupling constants depending on the moduli VEVs, 
the Higgs VEVs and the Wilson lines.

In Ref.~\cite{Abe:2012fj}, 
a semi-realistic pattern of the masses and mixing angles of the quarks and 
leptons 
was realized with certain values of them shown in Appendix \ref{sec:appa}. 
The top quark mass obtained there is slightly below the experimental data, 
that is, the value of the top Yukawa coupling $y_{33}^u$ is a bit small, 
because such a high-precision analysis was not required for the purpose 
in Ref.~\cite{Abe:2012fj} to study the flavor structure. 
However, large quantum corrections are required 
to realize the 126 GeV Higgs mass within the MSSM (or MSSM-like models) 
and the dominant contribution will come 
from the top Yukawa coupling $y_{33}^u$. 
The slightly small top Yukawa coupling induced by the ansatz adopted 
in Ref.~\cite{Abe:2012fj} 
will be a disadvantage to obtain a realistic model 
including the 126 GeV Higgs boson.

Therefore, for the purpose in this paper to study especially the Higgs mass, 
we adopt another ansatz a little different from the original one. 
Our new ansatz is the following; the VEVs of Higgs fields are 
\begin{eqnarray}
\tan\beta&\equiv& v_u/v_d = 15,\nonumber\\
\langle H_u^K\rangle &=& \left(~0.0,~~0.0,~~3.3,~~1.2,~~0.0,~~0.0~\right)
\times v_u\,\mathcal N_u,\nonumber\\
\langle H_d^K\rangle &=& \left(~0.0,~~0.1,~~5.9,~~5.9,~~0.0,~~0.1~\right)
\times v_d\,\mathcal N_d\,,
\label{eq:higpara}
\end{eqnarray}
where we use the normalization factors $\mathcal N_u=\sqrt{3.3^2+1.2^2}$ and 
$\mathcal N_d=\sqrt{2(0.1^2+5.9^2)}$. 
The VEVs of moduli are chosen as 
\begin{eqnarray}
\pi s &=& 6.0,\nonumber\\
\left(t_1,~t_2,~t_3\right) &=& \left(3.0,~1.0,~1.0\right)\times 2.8\times 10^{-8}, \nonumber\\
\left(u_1,~u_2,~u_3\right) &=& \left(4.4,~1.0,~1.0\right). 
\label{eq:modupara}
\end{eqnarray} 
The dilaton VEV $s$ leads to $4\pi/ g_a^2=24$ at the GUT scale 
$M_{\rm GUT}=2.0\times 10^{16}$, which is the unified value 
implemented by the MSSM\footnote{
Note again that all the exotics other than MSSM contents are 
assumed to be heavier than the compactification scale}. 
The VEVs $t_r$ ($r=1,2,3$) of K\"ahler moduli determine the 
compactification scale, which is fixed 
to $M_{\rm GUT}$, and the ratios between them are chosen to satisfy 
the SUSY condition (\ref{eq:arear}). 
As for the VEVs $u_r$ ($r=1,2,3$) of complex structure moduli, 
only the first $u_1$ is important 
because the flavor structure originates solely from the first 
torus $(T^2)_1$. 
The other $u_2$ and $u_3$ have no effect on the flavor structures 
and we set them as $u_2=u_3=1.0$ for simplicity.

The Wilson-lines, which mostly affect the Yukawa hierarchy, 
are selected as 
\begin{eqnarray}
\left(\zeta^{(1)}_C,~\zeta^{(1)}_{C'},~\zeta^{(1)}_L,~
\zeta^{(1)}_{R'},~\zeta^{(1)}_{R''}\right) = 
\left(0.0,-0.5i,-0.6i,~0.9i,~0.8i\right). 
\label{eq:wilpara}
\end{eqnarray} 
We have a degree of freedom to shift the wavefunctions of 
all the elements universally, that allows us to set $\zeta^{(1)}_C=0$ 
for simplicity without affecting the flavor structure. 
The above ansatz yields a semi-realistic pattern of the quark masses, 
the charged lepton masses and the CKM mixing angles \cite{Kobayashi:1973fv} 
at the electroweak (EW) scale $M_Z$ 
including contributions from the full 1-loop Renormalization Group Equation 
(RGE) which are shown in Table \ref{tab:ckm}. 
We obtain the larger top quark mass in spite of the smaller value 
of $\tan\beta$. 
The top Yukawa coupling $y_{33}^u=0.997$ becomes larger than 
$y_{33}^u=0.971$ obtained in Ref.~\cite{Abe:2012fj} at the EW scale. 
That causes an unnegligible effect because the dominant correction 
to the Higgs mass is proportional to $(y_{33}^u)^4$. 
Moreover, the smaller value of $\tan\beta$ has another advantage 
in the next section.

\begin{table}[t]
\begin{center}
\begin{tabular}{|c||c|c|} \hline
 & Sample values & Observed \\ \hline
$(m_u, m_c, m_t)$ & 
$(2.43 \times 10^{-3}, 0.431, 1.73 \times 10^2)$ & 
$(2.30 \times 10^{-3}, 1.28, 1.74 \times 10^2)$ 
\\ \hline
$(m_d, m_s, m_b)$ & 
$(4.59 \times 10^{-3}, 1.86\times 10^{-1}, 10.7)$ & 
$(4.8 \times 10^{-3}, 0.95 \times 10^{-1}, 4.18)$ 
\\ \hline
$(m_e, m_\mu, m_\tau)$ & 
$(1.53 \times 10^{-3}, 6.36 \times 10^{-2}, 5.11)$ & 
$(5.11 \times 10^{-4}, 1.06 \times 10^{-1}, 1.78)$ 
\\ \hline \hline 
$|V_{\rm CKM}|$ & 
\begin{minipage}{0.35\linewidth}
\begin{eqnarray} 
\left( 
\begin{array}{ccc}
0.987 & 0.161 & 0.00585 \\
0.159 & 0.982 & 0.0964 \\
0.0213 & 0.0942 & 0.995 
\end{array}
\right) 
\nonumber
\end{eqnarray} \\*[-20pt]
\end{minipage}
& 
\begin{minipage}{0.4\linewidth}
\begin{eqnarray} 
\left( 
\begin{array}{ccc}
0.97 & 0.23 & 0.0035 \\
0.23 & 0.97 & 0.041 \\
0.0087 & 0.040 & 1.0 
\end{array}
\right) 
\nonumber
\end{eqnarray} \\*[-20pt]
\end{minipage} \\ \hline
\end{tabular}
\end{center}
\caption{The sample theoretical values of the quark masses, the charged lepton masses and 
the CKM mixing angles, which are estimated at the EW scale through the full 1-loop RGE flows. 
The observed values are quoted from Ref.~\cite{Beringer:1900zz}. }
\label{tab:ckm}
\end{table}

Note that, as in Ref.~\cite{Abe:2012fj}, we assume the existence of 
supersymmetric Higgs mass term which could not be derived from the 10D 
SYM action. 
Non-perturbative effects and/or higher-dimensional operators might be able to 
generate the mass term,  
\begin{equation*}
\mu_{KL}H_u^KH_d^L, 
\end{equation*}
in the superpotential $W$ of the 4D effective theory. 
This $\mu_{KL}$ is a $(6\times6)$ matrix and, as mentioned previously, 
we assume that the five eigenvalues are as large as $M_{\rm GUT}$ 
and the last one, which will be identified with the so-called 
$\mu$-parameter of the MSSM, is much smaller than the other five 
comparable to the (low) SUSY breaking scale. 
The five linear combinations among the six generations of the pair of 
Higgs doublets ($H_u^K$, $H_d^K$) are decoupled from the MSSM 
at above the GUT scale and 
the last one pair is identified as the pair of MSSM Higgs doublets 
($H_u$, $H_d$) having the above mentioned $\mu$-term, $\mu H_u H_d$, 
in the superpotential\footnote{
We can also study with another scenario 
in which the five pairs of Higgs doublets other than the MSSM Higgs doublets 
are not decoupling from the MSSM at the GUT scale 
and would affect on the phenomenologies. We will study that elsewhere. }.

As for the neutrino sector, we can study it 
assuming the seesaw mechanism \cite{seesaw} with the following heavy Majorana mass term, 
\begin{equation*}
M^N=\begin{pmatrix}
0.1&1.9&0.0\\
1.9&0.3&3.1\\
0.0&3.1&1.4
\end{pmatrix}\times 10^{11}~ {\rm GeV},
\end{equation*}
which might be also generated by non-perturbative and/or higher-order effects. 
In this case, a semi-realistic pattern of the neutrino masses and the PMNS 
mixing angles \cite{Pontecorvo:1967fh}, 
as shown in Table \ref{tb:mns}, are produced at the same time 
as the quark and the charged lepton masses as well as the CKM mixing angles 
shown in Table~\ref{tab:ckm} are realized. 
\begin{table}[t]
\begin{center}
\begin{tabular}{|c||c|c|} \hline
 & Sample values & Observed 
\\ \hline
$(m_{\nu_1}, m_{\nu_2}, m_{\nu_3})$ & 
$(2.57 \times 10^{-17}, 1.11 \times 10^{-11}, 9.30 \times 10^{-11})$ & 
$< \ 2 \times 10^{-9}$ 
\\ \hline
$|m_{\nu_1}^2-m_{\nu_2}^2|$ & 
$1.24 \times 10^{-22}$ & 
$7.50 \times 10^{-23}$ 
\\ \hline
$|m_{\nu_1}^2-m_{\nu_3}^2|$ & 
$8.65 \times 10^{-21}$ & 
$2.32 \times 10^{-21}$ 
\\ \hline \hline 
$|V_{\rm PMNS}|$ & 
\begin{minipage}{0.3\linewidth}
\begin{eqnarray} 
\left( 
\begin{array}{ccc}
0.933 & 0.255 & 0.255 \\
0.354 & 0.779 & 0.518 \\
0.0668 & 0.574 & 0.816 
\end{array}
\right) 
\nonumber
\end{eqnarray} \\*[-20pt]
\end{minipage}
& 
\begin{minipage}{0.3\linewidth}
\begin{eqnarray} 
\left( 
\begin{array}{ccc}
0.82 & 0.55 & 0.16 \\
0.51 & 0.58 & 0.64 \\
0.26 & 0.61 & 0.75 
\end{array}
\right) 
\nonumber
\end{eqnarray} \\*[-20pt]
\end{minipage} \\ \hline
\end{tabular}
\end{center}
\caption{The sample theoretical values of the neutrino masses and 
the PMNS mixings, which are estimated at the EW scale through the full 1-loop RGE flows. 
The observed values are quoted from Ref.~\cite{Beringer:1900zz}.}
\label{tb:mns}
\end{table}

In this section, we have shown a sample spectrum of the SM particles 
with the ansatz (\ref{eq:higpara})-(\ref{eq:wilpara}). 
On this background, we will study the SUSY spectra and the Higgs boson mass 
in the next section. 
Finally, we emphasize that the observed mysterious hierarchical structure of 
quarks and leptons are successfully generated from the non-hierarchical 
input values of VEVs shown in Eqs.~(\ref{eq:higpara})-(\ref{eq:wilpara}), 
by virtue of the wavefunction localization of chiral zero-modes caused by 
magnetic fluxes.

\section{The Higgs boson mass and SUSY spectra}\label{sec:higgs}
We study the Higgs boson mass and SUSY spectra in the 4D effective SUGRA. 
We consider two types of SUSY breaking mediation mechanisms in this model, 
the moduli ($S$, $T_r$, $U_r$) and the anomaly 
(compensator $C$)~\cite{Randall:1998uk} mediations 
whose mixture is referred to as mirage mediation 
\cite{Choi:2005uz,Endo:2005uy}. 
The soft SUSY breaking parameters induced by them are calculated by the 
formulae given by Ref.~\cite{Choi:2005uz} and 
we estimate the spectra at the EW scale through the full 1-loop RGE flows 
with the MSSM contents numerically, 
and then we can study the Higgs boson mass.

\subsection{Soft parameters}
In this section, we assume a certain moduli stabilization mechanism works and 
it determines the VEVs of seven moduli scalar components, $s, t_r, u_r$, 
which is consistent with Eqs. (\ref{eq:higpara})-(\ref{eq:wilpara}) 
to realize the SM flavors. 
Our model has two types of SUSY breaking mediations and, 
although their contributions proportional to the mediator's $F$-terms 
will also be determined by the moduli stabilization 
as well as the SUSY breaking scenario, 
we study here treating the VEVs $F^m$ ($m=s,t_r,u_r$) and $F^C$ 
as free parameters representing the magnitude of the SUSY breaking 
mediated by each of the moduli and the anomaly respectively. 
The nonvanishing $F$-terms generate the soft SUSY breaking terms, and then, 
the soft parameters can be calculated and the spectrum is obtained 
depending on these parameters. 
In this way, we will be able to study the model concentrating on the 
Higgs boson mass and the SUSY spectrum 
without a concrete moduli stabilization scenario. 
Inversely speaking, the results of our analysis would probe the moduli 
stabilization as well as the SUSY breaking mechanisms behind our model.

As discussed above, we can calculate the soft parameters, 
which are the gaugino masses $M_a$, the scalar tri-linear couplings 
called $A$-terms $a_{IJK}$ and the soft scalar mass squares 
$(m^2_{\tilde Q})_I^J$, by using the following formulae~\cite{Choi:2005uz}, 
\begin{eqnarray}
M_a&=&\frac{F^s}{s+\bar s}+\frac{b_a}{16\pi^2} g_a^2 
\frac{F^C}{C}, \nonumber\\
a_{IJK}&=&y_{IJK} ( c_{Q_L}^m+c_{Q_R}^m+c_H^m) \frac{F^m}{\varphi_m+\bar\varphi_m}
+y_{IJK}F^m\partial_m {\rm ln} \lambda_{IJK} + 
(y_{LJK}\gamma_I^L+(I\leftrightarrow J,K))\frac{F^C}{C} ,\nonumber\\
({m^2_{\tilde Q}})_I^J&=& c_Q^m \left|\frac{F^m}{\varphi_m+\bar\varphi_m} \right|^2 \delta_I^J 
- \partial_m \gamma_I^J 
\left(\frac{F^m}{\varphi_m+\bar\varphi_m} \frac{\bar{F^C}}{\bar{C}}+{\rm h.c.}\right) 
+\frac{1}{4}\dot{\gamma}_I^J  \left| \frac{F^C}{C} \right|^2,
\label{eq:sofpara} 
\end{eqnarray} 
where $\gamma_I^J$ is the anomalous dimension and 
$\dot{\gamma}=\frac{\partial\gamma}{\partial \ln\,\mu/\Lambda}$. 
Note that, the index "$m$" is summed over the seven moduli supermultiplets 
$(S,~T_r,~U_r)$, 
and $\varphi_m=s, t_r, u_r$ represents the VEVs of their scalar components. 
The values of $c_Q^m$ ($Q=Q_L,~Q_R,~H$) derived from the K\"ahler metrics (\ref{eq:zzz}) 
are listed in Table \ref{tb:cweight}. 
\begin{table}
\begin{center}
\begin{tabular}{cccccccc}
\hline
     & $S$ & $T_1$ &  $T_2$  & $T_3$  & $U_1$  & $U_2$ & $U_3$ \\ \hline
$H$ &1/3&  -2/3 & 1/3 & 1/3 & -1/6 & -1/6 & -1/6 \\ 
$Q_L$ &1/3&  1/3 & -2/3 & 1/3 & -1/6 & -1/6 & 1/3 \\ 
$Q_R$ &1/3&  1/3 & 1/3 & -2/3 & -1/6 & 1/3 & -1/6 \\ \hline
\end{tabular}
\caption{The values of $c_Q^m$ appear in Eq.(\ref{eq:sofpara}).}
\label{tb:cweight}
\end{center}
\end{table} 
They are universal for their generations and 
determine how each moduli contributes to the soft masses 
of the left-handed $Q_L$, right-handed $Q_R$ and Higgs $H$ scalars. 

From Table \ref{tb:cweight}, 
we see that some of them give negative contributions. 
Negative contributions to the soft mass squares are disfavored in 
at least the left-handed and right-handed sectors to avoid tachyons 
and we expect that the total contributions given by the seven moduli should 
be positive\footnote{
Even if the anomaly mediation is included, we need the positive 
contribution somewhat 
because the pure anomaly mediation generally induces tachyonic sleptons. }. 
For instance, the five moduli other than $S$ and $T_1$ 
give the negative contributions to either the left or the right, or both. 
Thus one of them cannot solely contribute to the soft masses. 
We also find that the net contributions made by 
the three K\"ahler moduli $T_r$ vanish 
if the three contributions are equal, $F^{t_1}=F^{t_2}=F^{t_3}$. 
The gaugino masses are generated by dilaton $S$.  
We expect a naively  nonvanishing $F^s$ is required to some extent 
to generate gluino masses large enough satisfying the experimental 
lower bound. 
The moduli contributions to the $A$-terms are determined by the sums 
$c^m_{Q_L}+c^m_{Q_R}+c^m_H$, 
then we find $c^m_{Q_L}+c^m_{Q_R}+c^m_H\neq0$ for $m= S,U_1$ from 
Table \ref{tb:cweight} and the other moduli will not affect the $A$-terms. 
Moreover, we also find $c^S_{Q_L}+c^S_{Q_R}+c^S_H=1$ and that is equivalent 
to the so-called mirage condition \cite{Choi:2005uz}
if $F^{u_1}$ is negligible, 
which is required to avoid the SUSY flavor violations as shown later.

The moduli dependence of the soft terms (\ref{eq:sofpara}) is determined by 
the configuration of magnetic fluxes realizing the SM flavor structures. 
The K\"ahler modulus $T_i$ appears in the K\"ahler metrics of 
the chiral multiplets embedded in $\phi_i$ defined in Eq.~(\ref{eq:phii}) 
and it contains bosons behaving as vectors on only the $i$-th torus $(T^2)_i$ 
and their partners. 
For example, $Z^{(Q_L)}_{I\bar I}$ depends on the modulus $T_2$ as shown 
in Eq.~(\ref{eq:zzz}) 
because the left-handed matters $Q^I$ and $L^I$ are embedded into $\phi_2$. 
On the other hand, as for the complex structure moduli dependence, 
the modulus $U_i$ appears in the K\"ahler metrics if 
the corresponding multiplets feel nonvanishing magnetic fluxes on the 
$i$-th torus $(T^2)_i$. 
We see from Table \ref{tb:model336} that all the three sectors have 
nonvanishing fluxes on the first torus $(T^2)_1$, consequently, 
the complex structure modulus $U_1$ appears in their K\"ahler metrics. 
And also, the left-handed sector feels no magnetic fluxes on the third 
torus $(T^2)_3$, for example, 
and the modulus $U_3$ will not appear in its K\"ahler metric.

The modulus $U_1$ will receive the other severe constraint. 
The moduli dependence of all the K\"ahler metrics is universal 
for each generation involved in, while the modulus $U_1$ appears 
in the holomorphic Yukawa couplings (\ref{eq:lambda}) 
depending on the generation indices, because the three-generation 
structures are caused by the magnetic fluxes on the first torus $(T^2)_1$. 
That induces a severe restriction on the magnitude of the SUSY braking 
mediated by the complex structure modulus $U_1$ to suppress 
the SUSY flavor violations.

A phenomenological analysis of this model has in part been done 
in Ref.~\cite{Abe:2012fj} neglecting Yukawa matrix elements 
other than the most dominant one $y^{u,d,\nu,e}_{33}$. 
In this paper, we include the contributions from all of the Yukawa couplings 
to soft parameters at the GUT scale and their 1-loop RG effects, 
and we evaluate superparticle masses and SUSY flavor violations 
more precisely taking the latest experimental data into account. 
Furthermore, we estimate the Higgs boson mass, which has never estimated 
in this model, by calculating the 1-loop effective potential 
containing the top and bottom (s)quarks corrections 
\cite{Carena:1995wu}\footnote{
In Ref.~\cite{Carena:1995wu}, the leading log approximation 
was used, but we evaluate the self-coupling constant of 
the Higgs boson solving RGEs numerically for a better accuracy. }.

There are only the eight remaining parameters undetermined after realizing 
the semi-realistic SM sector. 
They are the VEVs $F^m$ ($m=s,t_r,u_r$) and $F^C$ of the $F$-terms 
of the moduli and compensator chiral supermultiplets, respectively. 
Since only the dilaton $S$ appears in the gauge kinetic functions and 
can give the masses to gauginos at the tree level, 
we refer to the normalized $F^s$ as the overall SUSY breaking scale,
\begin{align*}
M_{\rm SB}=\sqrt{K_{\bar{S}{S}}}F^s, 
\end{align*}     
and parameterize the other contributions by the following normalized ratios to $M_{\rm SB}$, 
\begin{align*}
R_r^T=\frac{\sqrt{K_{\bar{T}_rT_r}}F^{t_r}}{M_{\rm SB}},\ R_r^U=\frac{\sqrt{K_{\bar{U_r}U_r}}F^{t_r}}
{M_{\rm SB}},
\ R^C=\frac{1}{\ln M_p/m_{3/2}} \frac{F^C/C}{M_{\rm SB}}.
\end{align*} 
We show the results of numerical analyses varying these parameters 
in the following.

\subsection{Numerical results} 
\begin{figure}[t]
\centering
\includegraphics[width=0.45\linewidth]{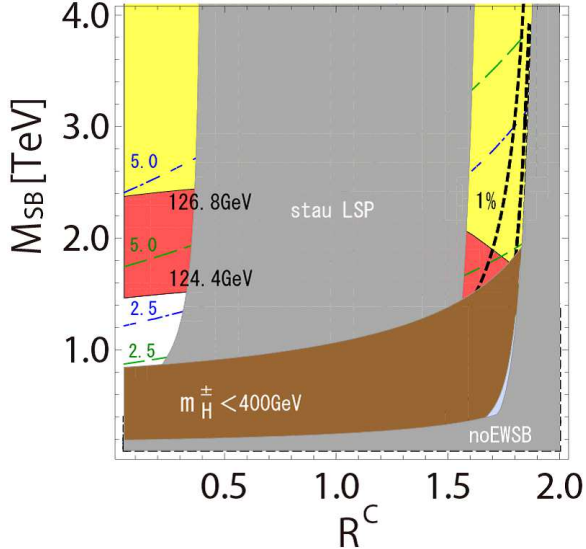} 
\caption{The contours of the Higgs boson masses and some experimentally 
relevant curves (regions) are drawn on the 
$(M_{\rm SB}~[{\rm TeV}], R^C)$-plane 
with $R^T_1=R^T_2=R^T_3=R^U_1=R^U_2=R^U_3=0$. 
In the red regions, the theoretical value of the Higgs boson mass 
resides in the allowed range, $124.4<m_h<126.8$ GeV. 
The Higgs boson is heavier than that in the yellow regions. } 
\label{fig:srun}
\end{figure}

First, we study $M_{\rm SB}$ dependence of the Higgs boson mass 
without the other moduli contributions, 
$R^T_1=R^T_2=R^T_3=R^U_1=R^U_2=R^U_3=0$. 
This situation corresponds to the simplest single modulus scenario and 
there remain two parameters, $M_{\rm SB}$ and $R^C$. 
We exhibit the contours of the Higgs boson mass and some 
observationally relevant curves (regions) on the 
$(M_{\rm SB}~[{\rm TeV}], R^C)$-plane and show that in Fig.~\ref{fig:srun}. 
In this figure, the theoretical value of the Higgs boson mass is inside 
the range of the experimental observations 
\cite{Aad:2012tfa,Chatrchyan:2012ufa}, 
$124.4<m_h<126.8$ GeV, in the red regions and exceeds that 
in the yellow regions. 
We find the 126 GeV Higgs boson can be realized with 
$M_{\rm SB}\sim 2.0$ TeV. In the gray regions, 
the EW symmetry breaking (EWSB) will not occur successfully or 
a stau becomes the Lightest Supersymmetric Particle (LSP), with which the 
pure MSSM cannot give any candidate for the dark matter. 
The green and blue dashed lines represent the masses (TeV) of the gluino 
and the lighter top squark respectively. 
The black dashed lines correspond to the degree of tuning the higgsino 
mass parameter ($\mu$-parameter) to obtain the observed $Z$-boson mass 
(radiatively), which is defined as $100/\Delta_\mu \,(\%)$ with 
\begin{equation*}
\Delta_\mu=\left|\frac{\partial \log m^2_Z}{\partial \log \mu^2}\right|.
\end{equation*}

The whole parameter space of ($M_{\rm SB}, R^C$) shown in Fig.~\ref{fig:srun} 
is free from the experimental constraints on various SUSY flavor violations 
estimated by evaluating the mass insertion parameters \cite{Misiak:1997ei}. 
This is mostly because here we set $R^U_1=0$, and the effect of $R^U_1\neq0$ 
will be shown later. 
We should remark that the lower bound on 
the charged Higgs boson mass is treated as $m_{H_\pm}>400$ GeV, 
because processes with charged Higgs boson exchange would contribute to 
$\Gamma (b\rightarrow s\gamma)$ and the charged Higgs boson 
lighter than $350$ GeV is disfavored\footnote{
The analysis with the mass insertion parameters cannot 
take into account this contribution, 
because the corresponding Feynman diagrams do not contain 
the soft parameters. }~\cite{Gambino:2001ew}. 
We exclude the region, where $m_{H_\pm}<400$ GeV, using brown shade.

Thus the suitable value of $M_{\rm SB}$ is about 2.0 TeV for the 126 GeV 
Higgs boson in this case, with which the SUSY spectra consistent 
with the various experimental results can be also obtained. 
Furthermore, the fine-tuning is relaxed better than $1\%$ with $R^C\sim 1.7$ and then 
almost all the soft parameters are unified at around TeV scale 
as pointed out in the TeV scale mirage mediation models \cite{Choi:2005hd} 
(See also Refs.~\cite{Abe:2007kf,Abe:2014kla}). 
Since the mirage unification at the TeV scale leads to the relatively light charged Higgs bosons 
as long as $|\mu|^2$ is small, consequently, some of the natural region 
$100/\Delta_{\mu} \gtrsim 1$ \% is covered by the brown shade, 
where the charged Higgs boson is lighter than 400 GeV 
because of the mirage unification, but some natural 
regions still remain allowed.

We remark that, with the small top Yukawa ansatz of Ref.~\cite{Abe:2012fj}, 
the 126 GeV Higgs mass requires the higher SUSY breaking scale than $2.0$ TeV, 
indeed, we could not find the allowed region with $M_{\rm SB}=2.0$ TeV. 
Furthermore, the small top Yukawa coupling will be accompanied with 
the large value of $\tan\beta$. 
In general, a large $\tan\beta$ induces the light charged Higgs boson 
and then broadens the excluded brown shade region in Fig.~\ref{fig:srun}, 
which will cover the natural region around $R^C\sim 1.7$. 
Thus the new Yukawa ansatz is more favored 
in order to realize the Higgs boson mass without the fine-tuning.

\begin{figure}[t]
\centering
\hfill
\includegraphics[width=0.45\linewidth]{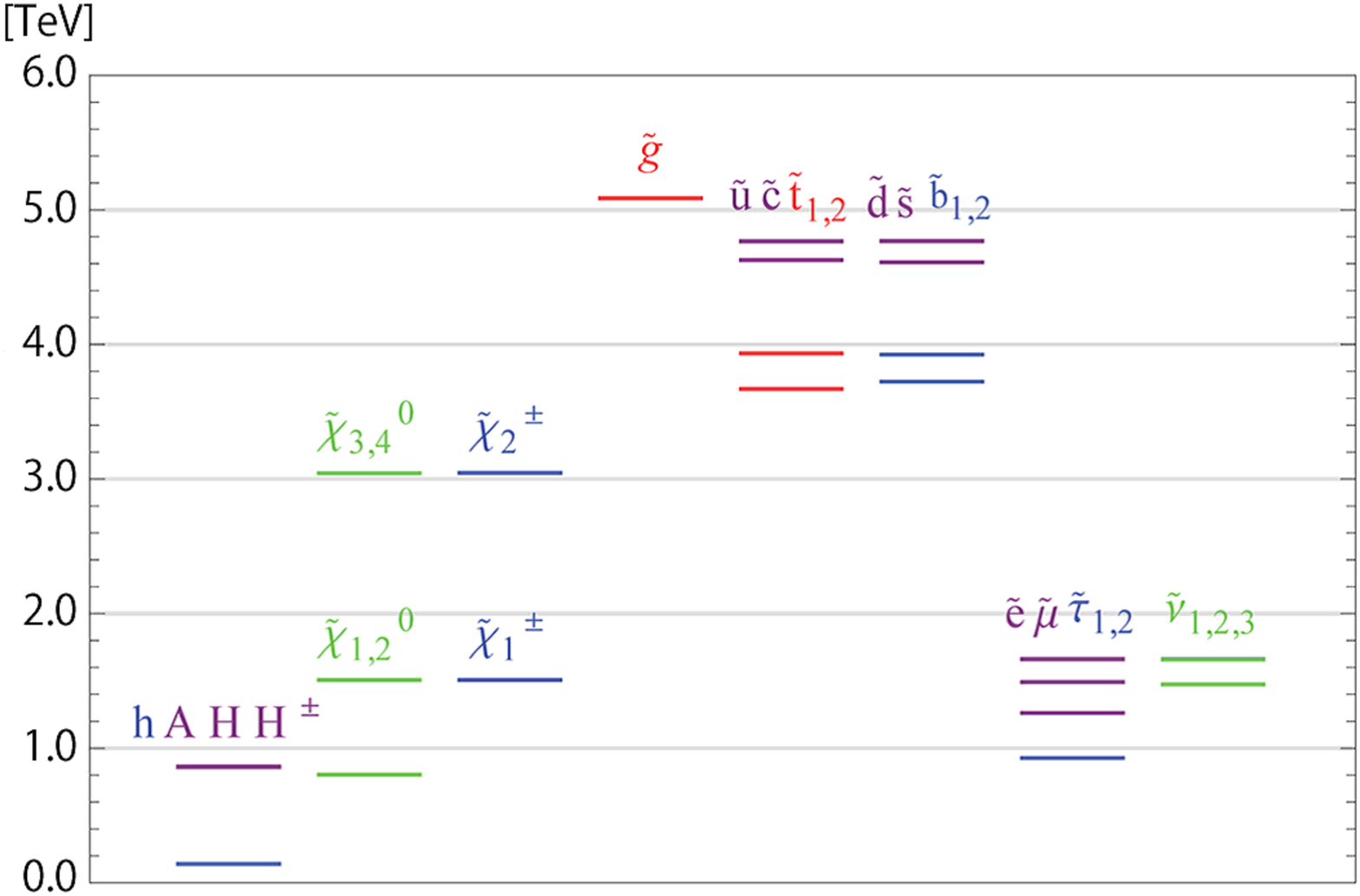} 
\hfill 
\includegraphics[width=0.45\linewidth]{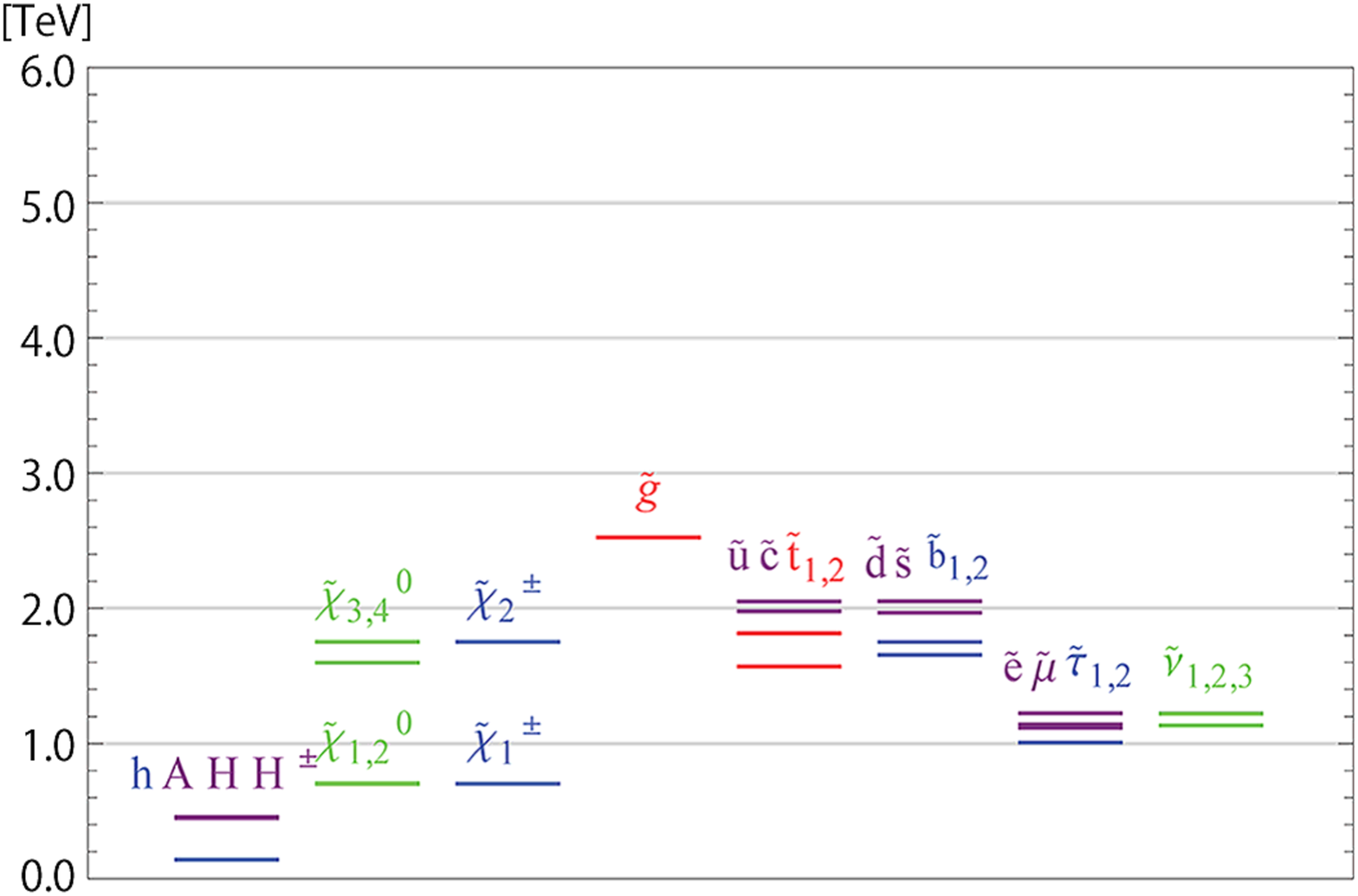}
\hfill 
\caption{The mass spectra of the MSSM particles with $M_{\rm SB}=1.8$ TeV and $R^C=0.1~(1.7)$ 
in the left (right) panel. } 
\label{fig:spec}
\end{figure}

We calculate the sample theoretical SUSY spectra given at the two different points 
in Fig.~\ref{fig:srun}. 
We show the spectrum derived from $(M_{\rm SB},~R^C)=(1.8~{\rm TeV},~0.1)$ 
in the left panel of Fig.~\ref{fig:spec} and 
the other one from $(1.8~{\rm TeV},~1.7)$ in the right. 
In the former (left) case, 
a CMSSM-like spectrum is realized 
with the 0.048 $\%$ tuning and the 125.4 GeV Higgs boson. 
In this case, the LSP is a bino-like neutralino and 
the colored  particles are heavier than the non-colored particles by sub-TeV. 
We can expect that sleptons or electroweakinos will be discovered earlier. 
In the latter (right) case, 
we get the 126.5 GeV Higgs boson with the relaxed tuning, 1.1 $\%$. 
The LSP is a higgsino-like neutralino, 
and most of the other sparticles are lighter than about 2.0 TeV 
and can be reached at the LHC in the near future. 
These two spectra carry the different LSP. 
We can further study this model with results of the cosmological observations 
if we consider the LSP dark matter, but we will not execute it here 
just leaving some comments. 
The bino dark matter scenario is being desperate as known. In the other case 
with a higgsino-like neutralino LSP, we can expect there are light colored 
SUSY particles and their masses are bounded by the Higgs observation 
because the theoretical value 
of the Higgs boson mass exceeds 
the observed one in the yellow regions in Fig.~\ref{fig:srun} 
(where the value of $\tan\beta$ is fixed). 
We might be able to verify this model by combination of the various experiments (at least with the parameters 
shown in Section \ref{sec:yukawa}).

\begin{figure}[t]
\centering
\includegraphics[width=0.45\linewidth]{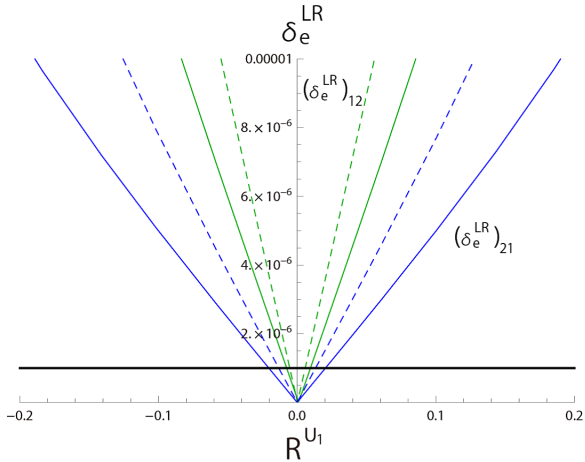} 
\caption{The mass insertion parameters $(\delta^{LR}_e)_{12}$ (green) and 
$(\delta^{LR}_e)_{21}$ (blue) as functions of $R^U_1$ 
with $R^C=0.1$ (solid) or $R^C=1.7$ (dashed). The black horizontal line represents 
$\delta_e^{LR}=10^{-6}$. } 
\label{muegamma}
\end{figure}

As mentioned above, the SUSY flavor violations are 
mostly depending on the complex structure modulus $U_1$ of the first torus 
$(T^2)_1$, and here we will study its effects. 
Within $M_{\rm SB}\le 1.0$ TeV, 
the modulus $U_1$ must not participate in the SUSY breaking mediation 
to suppress the dangerous SUSY flavor violations \cite{Abe:2012fj}, 
especially, concerning the process $\mu \rightarrow e\gamma$. 
We study the effect of $R^U_1\neq0$ more precisely with the above two 
sample spectra. 
We show the relevant mass insertion parameters as functions of $R^U_1$ 
in Fig.~\ref{muegamma} where $M_{\rm SB}=1.8 $ TeV and $R^C=0.1~(1.7)$ 
with the solid (dashed) lines, 
and the others are vanishing. 
They have the stringent constraints 
$\mathcal O ((\delta_e^{LR})_{12,21})\lesssim 10^{-6}$ 
(black horizontal line) given by Ref.~\cite{Arana-Catania:2013nha}. 
We see that the value of $R^U_1$ must be tiny, at least $|R^U_1|\lesssim 0.01$, 
even if the SUSY spectra are a little heavier to be consistent with the 
Higgs discovery. We can also satisfy all the other constraints on SUSY 
flavor violations from the flavor changing neutral current (FCNC) experiments 
easily with the tiny value of $|R^U_1|$. 
We adopt $R^U_1=0$ for simplicity in the following analysis\footnote{
The KKLT-type scenario \cite{Kachru:2003aw} of moduli stabilization, 
fixing all the complex structure moduli at a high-scale 
in a supersymmetric way, is one of the candidates for realizing 
$R^U_1\approx 0$.}.

Next, we study our model with the other moduli-mediated contributions. 
The moduli $T_2,~T_3,~U_2$ and $U_3$ of the other tori 
$(T^2)_2$ and $(T^2)_3$ than $(T^2)_1$ can solely give negative contributions to the squark and slepton mass squares 
as is seen from Table \ref{tb:cweight}. 
As mentioned before, the totally positive contributions are required 
in the left- and right-handed sectors to obtain non-tachyonic sparticles. 
(Furthermore, if the absolute values of the squared soft masses 
become so large in the Higgs sector, the fine-tuning problem can be serious.) 
We consider the case with $R^T_2=R^T_3$ and $R^U_2=R^U_3$ for simplicity. 
A combination of $U_2$ and $U_3$ behaves similarly to $T_1$ 
as long as we focus on the SUSY spectra, 
and we can choose $R^U_2=R^U_3=0$ without loss of generality.

\begin{figure}[t]
\centering
\hfill
\includegraphics[width=0.45\linewidth]{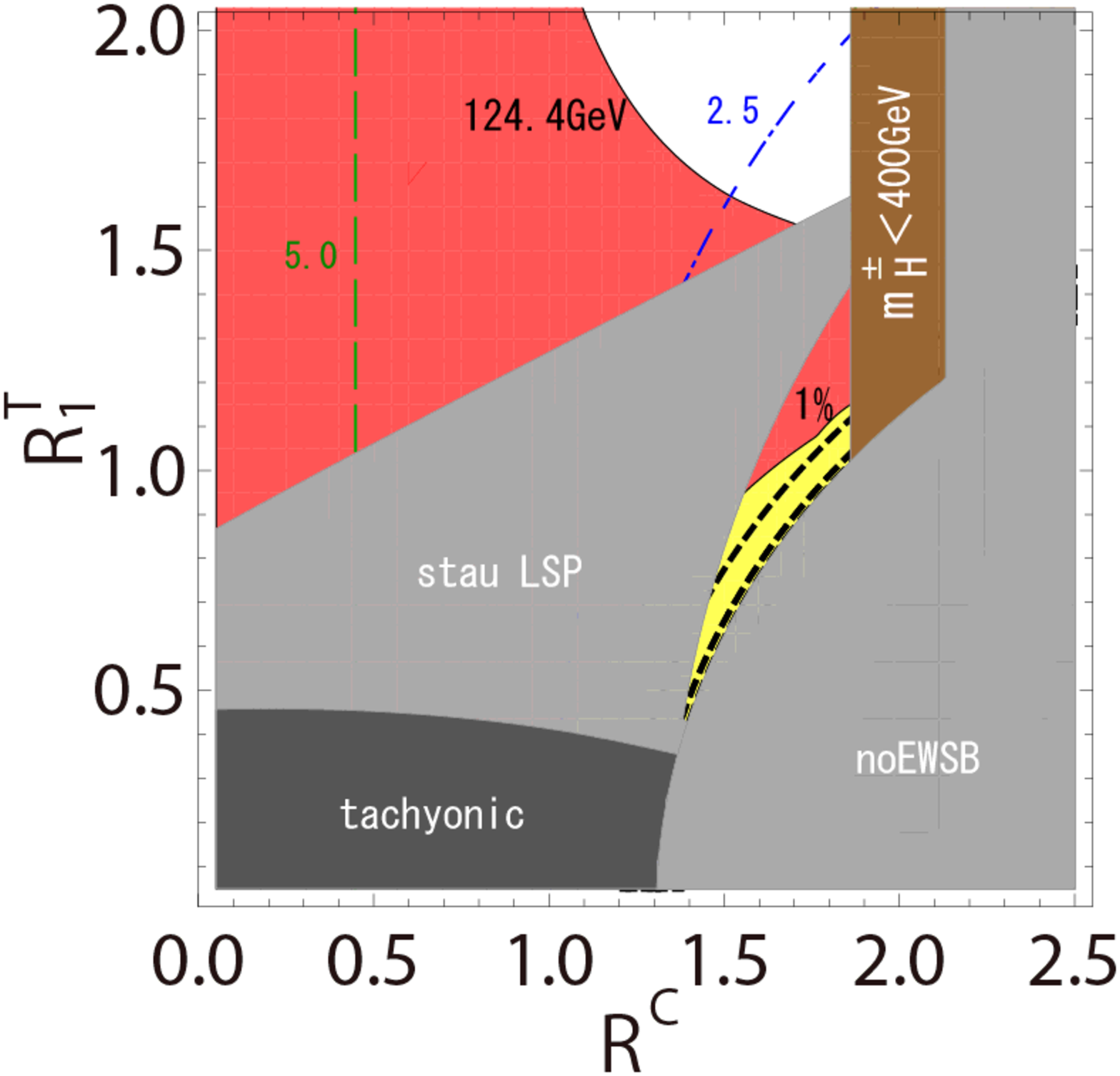} 
\hfill 
\includegraphics[width=0.45\linewidth]{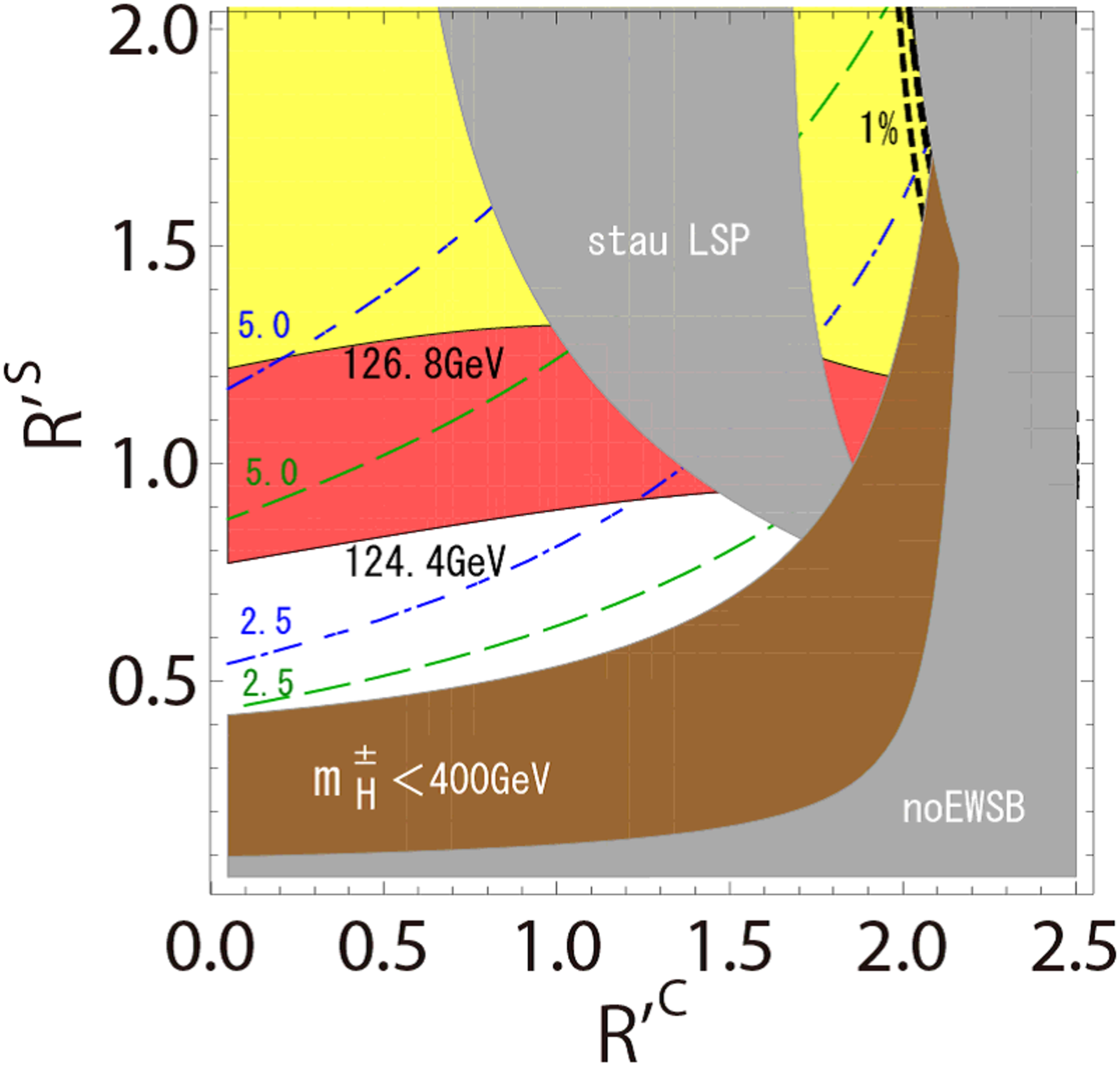}
\hfill 
\caption{The contours of the Higgs boson mass on the $(R^T_1, R^C)$-plane 
with $M_{\rm SB}=2.0$ TeV and $R^T_2=R^T_3=1$ in the left panel. 
We also represent various observational constraints similar to those in 
Fig.~\ref{fig:srun}. The similar drawing is shown on the $(R'^S, R'^C)$-plane 
with $M'_{\rm SB}=2.0$ TeV in the right panel. }
\label{fig:trun}
\end{figure}

We show again the contours of the Higgs boson mass and 
some observationally relevant curves (regions) on the $(R^T_1, R^C)$-plane 
with $M_{\rm SB}=2.0$ TeV and $R^T_2=R^T_3=1$ in the left panel of 
Fig.~\ref{fig:trun}. 
At the points with $R^T_1\sim1$, the net contributions 
from the three K\"ahler moduli $T_r$ almost vanish 
and dilaton $S$ dominates the SUSY breaking mediation. 
The negative contributions from $T_2$ and $T_3$ dominate below the points, 
$R^T_1<1$, while the positive one from $T_1$ becomes dominant above them, 
$R^T_1>1$. 
We can see the negative contributions comparable to that of the dilaton are dangerous, and 
\begin{equation*} 
\sum_{m\neq S} c_Q^m R^m \gtrsim 0, \qquad{\rm for}\qquad Q = Q_L,~Q_R, 
\end{equation*} 
is required in general cases.

Since some observationally excluded regions will shrink with $R^T_1\gtrsim 1$, 
we further study the $T_1$-dominant case in another analysis, 
where we renew some parameters as follows, 
\begin{equation*}
M'_{\rm SB}=\sqrt{K_{\bar{T_1} T_1}}F^{t_1}=2.0~ {\rm TeV},\qquad 
R'^S=\frac{\sqrt{K_{\bar S S}}F^s}{M'_{\rm SB}}, \qquad 
R'^C=\frac{1}{\ln M_p/m_{3/2}} \frac{F^C/C}{M'_{\rm SB}},
\end{equation*} 
and the other $F$-terms are vanishing. 
Similarly we show the contours of the Higgs boson mass and some experimentally 
relevant curves (regions) on the ($R'^S, R'^C$)-plane 
in the right panel of Fig.~\ref{fig:trun}. 
Compared with Fig.~\ref{fig:srun}, it is obvious that the positive 
contributions from $T^1$ 
to the sfermions are favored. 
However, we can also see the dilaton contribution is indispensable 
to a certain extent even if with the others contribute. 
The main reason of that is the large gluino mass is required, 
especially in the RGE flows, to yield the successful EWSB 
and that can be generated by only the dilaton and anomaly, 
because, as explained at the end of Section \ref{sec:review}, 
the gauge kinetic functions are functions of only the dilaton 
in 10D SYM theory.

\section{Conclusions and Discussions}\label{sec:conc}
We have studied the Higgs boson mass and precise SUSY spectra 
of the particle physics model derived from the 10D SYM theory compactified 
on the magnetized tori. 
This model was proposed in Ref.~\cite{Abe:2012fj}, where 
the magnetic fluxes in the extra dimensional space originate the complicated 
flavor structures of the SM 
and the magnitude of the SUSY flavor violations was estimated with rough 
approximations to check consistency. 
In this paper, we are focusing on the 126 GeV Higgs boson mass. 
For such a purpose, 
we have first done a minor improvement of the Yukawa structures 
to enhance the top Yukawa coupling $y^u_{33}$ 
realizing the masses and mixing angles of the quarks and leptons. 
The enhanced top Yukawa coupling $y^u_{33}$ can be an advantage to generate 
the large quantum corrections 
required to realize the 126 GeV Higgs boson within low scale 
SUSY breaking scenarios.

On the improved background, we have studied the model using the SUSY breaking 
parameters $F^m$ and $F^C$ 
of the moduli and compensator chiral multiplets, respectively. 
We have estimated the Higgs boson mass in the simplest case by a varying 
the overall SUSY breaking scale $M_{\rm SB}$, 
and obtained the 126 GeV Higgs boson mass when the overall scale 
is about 2.0 TeV. 
In this case, the fine-tuning can be relaxed 
if there is a comparable contribution from the anomaly mediation 
with which the TeV scale mirage scenario \cite{Choi:2005hd} is realized. 
We have shown the two sample SUSY spectra allowed by the 
experimental constraints. 
One is a moduli dominated scenario and the other corresponds to the TeV scale 
mirage scenario. 
Both spectra will be reached at future experiments. 
They have some significant differences 
in the mass scales of the colored particles and the LSP constituent. 
These differences provide a motivation to study the model 
from the perspective of not only the high-energy experiments 
but also the cosmological observations elsewhere in a separate work. 
In particular, results from the combined studies might be able to 
verify this model in the near future.

We notice that the unimproved original flux configuration given in 
Ref.~\cite{Abe:2012fj} requires the higher SUSY breaking scale, 
and the wide regions of the parameter space including the natural regions 
$100/\Delta_\mu \gtrsim 1$ \% will be excluded because of the light 
charged Higgs boson possibly induced by the large value of $\tan\beta$. 
Our new ansatz of the Yukawa matrices is more favored clearly 
from these points of view.

With the two sample SUSY spectra, we have estimated the magnitude of 
SUSY flavor violations 
without the approximations adopted in Ref.~\cite{Abe:2012fj} and compared 
with the latest experimental data. 
The most stringent bound is still coming from the process 
$\mu\rightarrow e\gamma$ 
and it requires that the complex structure modulus $U_1$ of 
the first torus $(T^2)_1$, 
where the three-generation structures are solely caused, 
should not mediate the SUSY breaking contributions, 
$|R^U_1|\lesssim0.01$. 
In other words, $U_1$ modulus should be quite heavy enough to decouple from the MSSM sector 
in the low energy effective field theory. 
In such a case, all the other SUSY FCNC constraints are easily satisfied.

We have also studied the other moduli dependence of the MSSM spectra. 
There are some phenomenological constraints coming from other than 
the observed 126 GeV Higgs mass, 
e.g, the FCNCs, the successful (radiative) EWSB (the observed $Z$-boson mass), 
no tachyons (color and charge breaking minima) and the (LSP) dark matter 
candidate, etc. 
They restrict the dynamics of the moduli, and we find the following 
indications. 
First, dilaton $S$ has to contribute to the SUSY breaking mediations 
to some extent even if the other contributions can be expected or not, 
because the successful EWSB requires the large gluino mass and 
the dilaton contribution is indispensable for that. 
Second, negative contributions to the squark and slepton mass squares 
comparable to that from the dilaton 
should be forbidden to avoid tachyonic particles.

We see that the various experimental results restrict the VEVs $s,t_r,u_r$ 
and moduli $F$-terms $F^s,F^{t_r},F^{u_r}$, those has their own geometrical 
meanings, as well as the $F$-term of chiral compensator $F^C$. 
The further high-energy experiments will be able to prove the SUSY breaking 
mediation mechanisms in our model and, furthermore, the moduli stabilization 
and SUSY breaking mechanisms behind it. 
Based on the results obtained in this paper, 
we will study a concrete moduli stabilization scenario  
including a SUSY breaking sector elsewhere. 
Since the recent cosmological observations is quite promoted, 
it would also be attractive to study about cosmological issues 
based on our model as mentioned in the previous section.

We can also consider some other extensions of this model including 
the gauge mediations. 
Our SUSY spectra given in this paper can be deflected in such a 
extended models. 
Although the additional gauge mediation is not expected to be 
a promising advantage from the naturalness perspective \cite{Abe:2014kla}, 
we might be able to expect drastic changes of the spectra 
and be inspired to go on to more various phenomenologies.

\subsection*{Acknowledgement}
H.A. was supported in part by the Grant-in-Aid for Scientific Research No.~25800158 
from the Ministry of Education, Culture, Sports, Science 
and Technology (MEXT) in Japan. 
J.K. was supported in part by the Grant for Excellent Graduate
Schools from the MEXT in Japan. 
K.S. was supported in part by a Grant-in-Aid for JSPS Fellows 
No.~25$\cdot$4968 and a Grant for Excellent Graduate
Schools from the MEXT in Japan. 

\appendix

\section{Yukawa matrices of the previous work}
\label{sec:appa}
We show, for a reference, the original input parameters adopted 
in Ref.~\cite{Abe:2012fj} below,
\begin{eqnarray*}
\tan\beta&=&25,\\
\langle H_u^K\rangle &=& \left(~0.0,~~0.0,~~2.7,~~1.3,~~0.0,~~0.0~\right)
\times v_u\,\mathcal N_u,\\
\langle H_d^K\rangle &=& \left(~0.0,~~0.1,~~5.8,~~5.8,~~0.0,~~0.1~\right)
\times v_d\,\mathcal N_d\,,\\
\pi s &=& 6.0,\\
\left(t_1,~t_2,~t_3\right) &=& \left(3.0,~1.0,~1.0\right)\times 2.8\times 10^{-8}, \\
\left(u_1,~u_2,~u_3\right) &=& \left(4.1,~1.0,~1.0\right),\\
\left(\zeta^{(1)}_C,~\zeta^{(1)}_{C'},~\zeta^{(1)}_L,~
\zeta^{(1)}_{R'},~\zeta^{(1)}_{R''}\right) &=& 
\left(0.0,~0.3i,-1.0i,~1.9i,~1.4i\right),
\end{eqnarray*}
where the Majorana neutrino masses were assumed as 
\begin{equation*}
M^N=\begin{pmatrix}
1.1&1.3&0\\
1.3&0&3.2\\
0&3.2&1.8
\end{pmatrix}\times 10^{12} ~{\rm GeV}.
\end{equation*}
From these input parameters, the theoretical values of quark and lepton 
masses and mixing angles shown in Table \ref{tab:ckmapp} and \ref{tb:mnsapp} 
are obtained, which are compared with those shown in Table 
\ref{tab:ckm} and \ref{tb:mns} derived from the improved parameters proposed 
in this paper. 
\begin{table}[htb]
\begin{center}
\begin{tabular}{|c||c|c|} \hline
 & Sample values & Observed \\ \hline
$(m_u, m_c, m_t)$ & 
$(3.1 \times 10^{-3}, 1.01, 1.70 \times 10^2)$ & 
$(2.30 \times 10^{-3}, 1.28, 1.74 \times 10^2)$ 
\\ \hline
$(m_d, m_s, m_b)$ & 
$(2.8 \times 10^{-3}, 1.48\times 10^{-1}, 6.46)$ & 
$(4.8 \times 10^{-3}, 0.95 \times 10^{-1}, 4.18)$ 
\\ \hline
$(m_e, m_\mu, m_\tau)$ & 
$(4.68 \times 10^{-4}, 5.76 \times 10^{-2}, 3.31)$ & 
$(5.11 \times 10^{-4}, 1.06 \times 10^{-1}, 1.78)$ 
\\ \hline \hline 
$|V_{\rm CKM}|$ & 
\begin{minipage}{0.35\linewidth}
\begin{eqnarray} 
\left( 
\begin{array}{ccc}
0.98 & 0.21 & 0.0023 \\
0.21 & 0.98 & 0.041 \\
0.011 & 0.040 & 1.0 
\end{array}
\right) 
\nonumber
\end{eqnarray} \\*[-20pt]
\end{minipage}
& 
\begin{minipage}{0.4\linewidth}
\begin{eqnarray} 
\left( 
\begin{array}{ccc}
0.97 & 0.23 & 0.0035 \\
0.23 & 0.97 & 0.041 \\
0.0087 & 0.040 & 1.0 
\end{array}
\right) 
\nonumber
\end{eqnarray} \\*[-20pt]
\end{minipage} \\ \hline
\end{tabular}
\end{center}
\caption{The sample theoretical values of the quark and charged lepton masses and 
the CKM mixing angles at the EW scale through the 1-loop RGE flows, 
those are derived from the input parameters given in Ref.~\cite{Abe:2012fj}. } 
\label{tab:ckmapp}
\end{table}
\begin{table}[ht]
\begin{center}
\begin{tabular}{|c||c|c|} \hline
 & Sample values & Observed 
\\ \hline
$(m_{\nu_1}, m_{\nu_2}, m_{\nu_3})$ & 
$(3.6 \times 10^{-19}, 8.8 \times 10^{-12}, 2.7 \times 10^{-11})$ & 
$< \ 2 \times 10^{-9}$ 
\\ \hline
$|m_{\nu_1}^2-m_{\nu_2}^2|$ & 
$7.67 \times 10^{-23}$ & 
$7.50 \times 10^{-23}$ 
\\ \hline
$|m_{\nu_1}^2-m_{\nu_3}^2|$ & 
$7.12 \times 10^{-22}$ & 
$2.32 \times 10^{-21}$ 
\\ \hline \hline 
$|V_{\rm PMNS}|$ & 
\begin{minipage}{0.3\linewidth}
\begin{eqnarray} 
\left( 
\begin{array}{ccc}
0.85 & 0.46 & 0.25 \\
0.50 & 0.59 & 0.63 \\
0.15 & 0.66 & 0.73 
\end{array}
\right) 
\nonumber
\end{eqnarray} \\*[-20pt]
\end{minipage}
& 
\begin{minipage}{0.3\linewidth}
\begin{eqnarray} 
\left( 
\begin{array}{ccc}
0.82 & 0.55 & 0.16 \\
0.51 & 0.58 & 0.64 \\
0.26 & 0.61 & 0.75 
\end{array}
\right) 
\nonumber
\end{eqnarray} \\*[-20pt]
\end{minipage} \\ \hline
\end{tabular}
\end{center}
\caption{The sample theoretical values of the neutrino masses and 
the MNS mixings at the EW scale through the 1-loop RGE flows, 
those are derived from the input parameters given in Ref.~\cite{Abe:2012fj}. }
\label{tb:mnsapp}
\end{table}


\begin{thebibliography}{99}



\bibitem{Aad:2012tfa}
  G.~Aad {\it et al.}  [ATLAS Collaboration],
  Phys.\ Lett.\ B {\bf 716} (2012) 1
  [arXiv:1207.7214 [hep-ex]].
  
  
\bibitem{Chatrchyan:2012ufa}
  S.~Chatrchyan {\it et al.}  [CMS Collaboration],
  Phys.\ Lett.\ B {\bf 716} (2012) 30
  [arXiv:1207.7235 [hep-ex]].


\bibitem{Bachas:1995ik}
  C.~Bachas,
  hep-th/9503030;
  C.~Angelantonj, I.~Antoniadis, E.~Dudas and A.~Sagnotti,
  Phys.\ Lett.\ B {\bf 489} (2000) 223
  [hep-th/0007090].

\bibitem{Cremades:2004wa}
  D.~Cremades, L.~E.~Ibanez and F.~Marchesano,
  JHEP {\bf 0405} (2004) 079  [hep-th/0404229].

\bibitem{Abe:2012ya} 
  H.~Abe, T.~Kobayashi, H.~Ohki and K.~Sumita,
  Nucl.\ Phys.\ B {\bf 863}, 1 (2012)
  [arXiv:1204.5327 [hep-th]].
  
\bibitem{Abe:2012fj}
  H.~Abe, T.~Kobayashi, H.~Ohki, A.~Oikawa and K.~Sumita,
  Nucl.\ Phys.\ B {\bf 870} (2013) 30
  [arXiv:1211.4317 [hep-ph]].



\bibitem{Marcus:1983wb}
  N.~Marcus, A.~Sagnotti and W.~Siegel,
  Nucl.\ Phys.\ B {\bf 224} (1983) 159; 
  N.~Arkani-Hamed, T.~Gregoire and J.~G.~Wacker,
  JHEP {\bf 0203} (2002) 055
  [hep-th/0101233].

\bibitem{Green:1984sg}
  M.~B.~Green and J.~H.~Schwarz,
  Phys.\ Lett.\ B {\bf 149} (1984) 117.

\bibitem{Abe:2013bba} 
  H.~Abe, T.~Kobayashi, H.~Ohki, K.~Sumita and Y.~Tatsuta,
  arXiv:1307.1831 [hep-th].

\bibitem{Kobayashi:1973fv}
  M.~Kobayashi and T.~Maskawa,
  Prog.\ Theor.\ Phys.\  {\bf 49} (1973) 652.

\bibitem{Beringer:1900zz}
  J.~Beringer {\it et al.}  [Particle Data Group Collaboration],
  Phys.\ Rev.\ D {\bf 86} (2012) 010001.

\bibitem{seesaw}
  T. Yanagida, in Proceedings of the Workshop on the Unified Theory 
and Baryon Number in the Universe, Tsukuba, Japan, 1979, 
eds. O. Sawada and A. Sugamoto, KEK repot KEK-79-13, p.95, and 
 Prog.\ Theor.\ Phys.\  {\bf 64} (1980) 1103; \\
  M. Gell-Mann, P. Ramond and R. Slansky, in Supergravity, 
North Holland, Amsterdam, 1979, eds. P. van Nieuwenhuizen and D. Z. Freedman, 
Print-80-0576 (CERN), p.315. 

\bibitem{Pontecorvo:1967fh}
  B.~Pontecorvo,
  Sov.\ Phys.\ JETP {\bf 26} (1968) 984
   [Zh.\ Eksp.\ Teor.\ Fiz.\  {\bf 53} (1967) 1717]; 
  Z.~Maki, M.~Nakagawa and S.~Sakata,
  Prog.\ Theor.\ Phys.\  {\bf 28} (1962) 870.


\bibitem{Randall:1998uk}
  L.~Randall and R.~Sundrum,
  Nucl.\ Phys.\ B {\bf 557} (1999) 79
  [hep-th/9810155]; 
  G.~F.~Giudice, M.~A.~Luty, H.~Murayama and R.~Rattazzi,
  JHEP {\bf 9812} (1998) 027
  [hep-ph/9810442].

\bibitem{Choi:2005uz}
  K.~Choi, K.~S.~Jeong and K.~-i.~Okumura,
  JHEP {\bf 0509} (2005) 039
  [hep-ph/0504037]. 

\bibitem{Endo:2005uy}
  M.~Endo, M.~Yamaguchi and K.~Yoshioka,
  Phys.\ Rev.\ D {\bf 72} (2005) 015004
  [hep-ph/0504036].

\bibitem{Carena:1995wu}
  M.~S.~Carena, M.~Quiros and C.~E.~M.~Wagner,
  Nucl.\ Phys.\ B {\bf 461} (1996) 407
  [hep-ph/9508343].

\bibitem{Misiak:1997ei} 
  M.~Misiak, S.~Pokorski and J.~Rosiek,
  Adv.\ Ser.\ Direct.\ High Energy Phys.\  {\bf 15}, 795 (1998)
  [hep-ph/9703442].



\bibitem{Gambino:2001ew} 
  P.~Gambino and M.~Misiak,
  Nucl.\ Phys.\ B {\bf 611}, 338 (2001)
  [hep-ph/0104034].

\bibitem{Choi:2005hd}
  K.~Choi, K.~S.~Jeong, T.~Kobayashi and K.~i.~Okumura,
  Phys.\ Lett.\ B {\bf 633} (2006) 355
  [hep-ph/0508029]; 
  R.~Kitano and Y.~Nomura,
  Phys.\ Lett.\ B {\bf 631} (2005) 58
  [hep-ph/0509039]; 
  K.~Choi, K.~S.~Jeong, T.~Kobayashi and K.~i.~Okumura,
  Phys.\ Rev.\ D {\bf 75} (2007) 095012
  [hep-ph/0612258].

\bibitem{Abe:2007kf}
  H.~Abe, T.~Kobayashi and Y.~Omura,
  Phys.\ Rev.\ D {\bf 76} (2007) 015002
  [hep-ph/0703044 [HEP-PH]]; 
  H.~Abe, J.~Kawamura and H.~Otsuka,
  PTEP {\bf 2013} (2013) 013B02
  [arXiv:1208.5328 [hep-ph]].

\bibitem{Abe:2014kla}
  H.~Abe and J.~Kawamura,
  arXiv:1405.0779 [hep-ph].


\bibitem{Arana-Catania:2013nha} 
  M.~Arana-Catania, S.~Heinemeyer and M.~J.~Herrero,
  Phys.\ Rev.\ D {\bf 88}, 015026 (2013)
  [arXiv:1304.2783 [hep-ph]].
  

\bibitem{Kachru:2003aw}
  S.~Kachru, R.~Kallosh, A.~D.~Linde and S.~P.~Trivedi,
  Phys.\ Rev.\ D {\bf 68} (2003) 046005
  [hep-th/0301240].






\end{thebibliography}
\end{document}